\documentclass[english,twocolumn,prx,amssymb,amsmath,superscriptaddress]{revtex4}
\usepackage[latin9]{inputenc}
\usepackage{graphicx,color}
\usepackage[bookmarks]{hyperref}
\usepackage{babel}
\usepackage{braket}

\newcommand{\be}{\begin{equation}}
\newcommand{\ee}{\end{equation}}
\newcommand{\bea}{\begin{eqnarray}}
\newcommand{\eea}{\end{eqnarray}}

\definecolor{mygreen}{rgb}{0,0.5,0}
\definecolor{myblue}{rgb}{0,0,0.75}
\definecolor{mymagenta}{cmyk}{0,1,0,0.12}

\usepackage{umoline}

\begin{document}

\title{Connecting dynamical quantum phase transitions and topological steady-state transitions
by tuning the energy gap}

\author{Pei Wang}
\affiliation{Department of Physics, Zhejiang Normal University, Jinhua 321004, China}
\author{Gao Xianlong}
\affiliation{Department of Physics, Zhejiang Normal University, Jinhua 321004, China}

\begin{abstract}
Considerable theoretical and experimental efforts have been devoted to the quench
dynamics, in particular, the dynamical quantum phase transition (DQPT) and the
steady-state transition. These developments have motivated us to
study the quench dynamics of the topological systems, from which we find the
connection between these two transitions, that is, the DQPT, accompanied by
a nonanalytic behavior as a function of time, always merges into a
steady-state transition signaled by the nonanalyticity
of observables in the steady limit. As the
characteristic time of the DQPT diverges, it exhibits universal scaling behavior,
which is related to the scaling behavior at the corresponding steady-state transition.
\end{abstract}
\maketitle

\date{\today}


\section{Introduction}
\label{sec:int}
Isolated quantum many-body systems can nowadays be realized in
quantum-optical systems such as ultra-cold atoms or trapped ions which
has opened up the perspective to experimentally study properties beyond
the thermodynamic equilibrium paradigm. This includes the observation of
genuine nonequilibrium phenonema such as many-body localization~\cite{Schreiber:2015,Smith:2016,Choi:2016},
quantum time crystals~\cite{Zhang:2017,Choi:2017}, or particle-antiparticle
production in the Schwinger model~\cite{Martinez:2016}. It remains, however,
a major challenge to identify universal properties in these diverse dynamical
phenomena on general grounds. Considerable effects have been made to the formulation of various notions
of nonequilibrium phase transitions~\cite{Yuzbashyan2006,Diehl2008,Barmettler2009gd,
Eckstein2009wj,Diehl2010,Sciolla2010jb,Garrahan2010xw,Mitra2012,Heyl2013a,Wang2016,Wang2016a}
which are seen as promising attempts to extend elementary equilibrium concepts such
as scaling and universality to the nonequilibrium regime. Among these
notions there is the concept of a steady-state transition, which is signaled by
a nonanalytic change of physical properties as a function of a parameter of the
nonequilibrium protocol in the asymptotic long-time state of the system~\cite{Diehl2008,Diehl2010,Sciolla2010jb}.
An example is the universal logarithmic divergence of the Hall conductance
in the steady state of topological insulators after a quench~\cite{Wang2016,Wang2016a}.
Another important concept is that of dynamical quantum phase transitions (DQPTs)~\cite{Heyl2013a},
recently observed experimentally~\cite{Flaeschner2016,Jurcevic2016},
which occur on transient and intermediate time scales accompanied by
a nonanalytic behavior as a function of time instead of a conventional order
parameter. In the context of such developments, it is then a natural question to ask,
whether and how these two notions of phase transitions are connected.

\begin{figure}
\centering
\includegraphics[width = \linewidth]{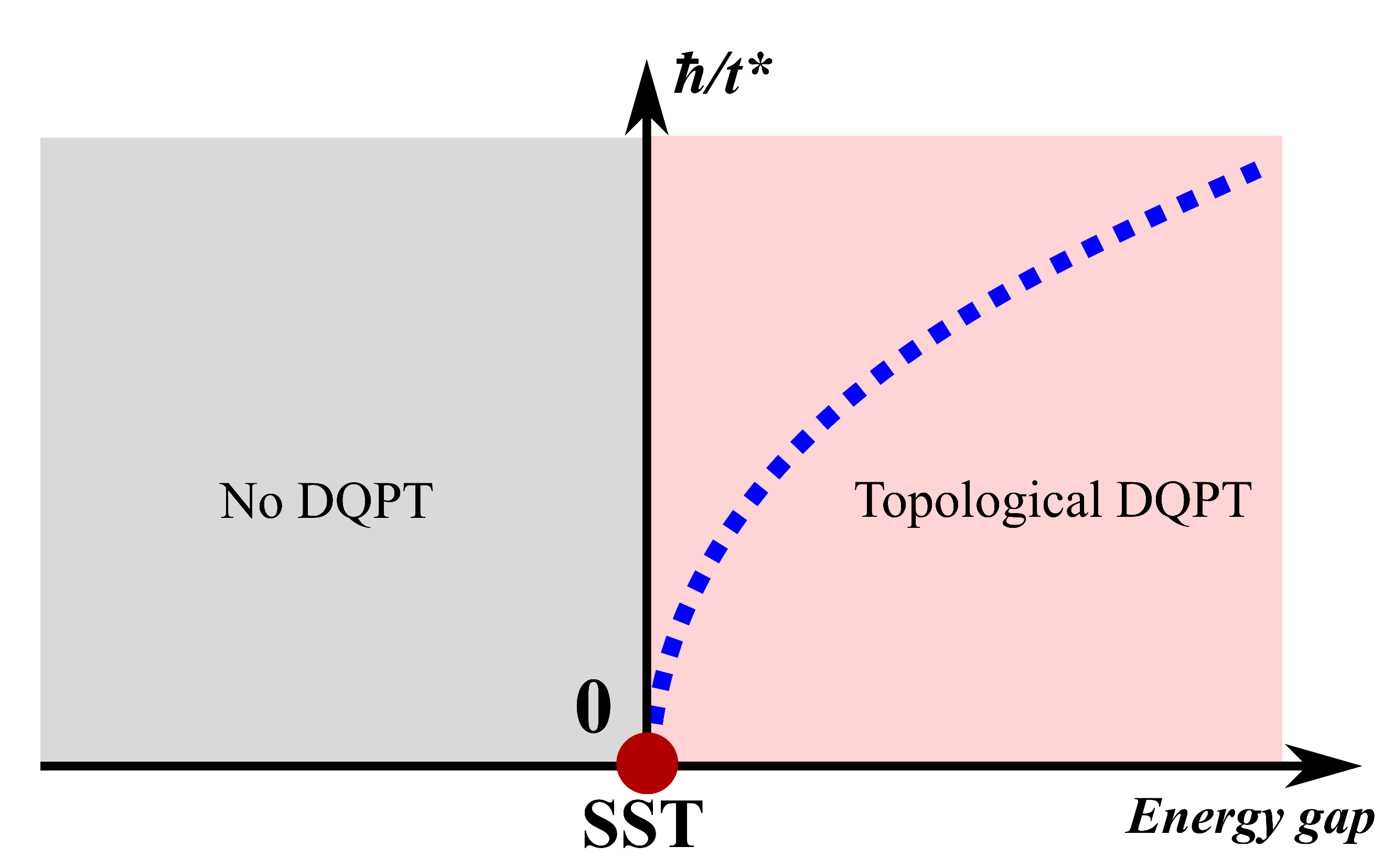}
\caption{The inverse of the characteristic
time $t^\ast$ associated with DQPTs is plotted as the energy gap $m$ towards the steady-state
transition occurring at $m = 0$ where the gap closes and reopens.
The DQPTs in the pink area is controlled by the steady-state transition (SST)
at the brown spot. DQPTs do not happen in the grey area.}
\label{fig:primary}
\end{figure}

The DQPT and the steady-state transition under the symmetry-breaking picture
have been recently connected in the long-range-interacting Ising chain~\cite{Zunkovic2016}
by relating the singularities of Loschmidt echo to the zeros of local order parameters.
In this work, we study the connection between the
DQPT and the steady-state transition in a topological system where the local order parameters are absent.

Vajna and D\'{o}ra~\cite{dora15} related the DQPTs to
the topological invariants in a two-band topological insulator.
They proved that the DQPTs happen whenever the ground-state topological
numbers of the initial and the post-quench Hamiltonians differ from each other.
This relation was then generalized to the multi-band models~\cite{Huang}.
On the other hand, the continuous but nonanalytic behavior in the Hall conductance
addresses a nonequilibrium steady-state transition as the energy
gap of the post-quench Hamiltonian closes~\cite{Wang2016,Wang2016a}.
In this paper, we employ the approach of singularity analysis developed in Refs.~\cite{Wang2016,Wang2016a}
to obtain a quantitative relation between the characteristic time scale $t^\ast$ associated with DQPTs
and the energy gap associated with the steady-state transition. Qualitatively speaking,
the divergence of $t^\ast$ must be accompanied by the gap closing and then a steady-state transition
(see Fig.~\ref{fig:primary} for the schematic diagram), which agrees to what was
found in the long-range-interacting Ising model~\cite{Zunkovic2016}.

More important, we find that the steady-state transition controls the DQPTs
in its vicinity, illustrated in the scaling of the dynamical free energy and the dynamics of vortices.
By using the singularity analysis, we obtain in the first time the asymptotic behavior
of the dynamical free energy at the DQPTs as the characteristic time diverges.
The dynamics of vortices was observed in a recent experiment~\cite{Flaeschner2016}.
Its relation to the topological numbers was then analyzed~\cite{Flaeschner2016,Wang17,Heyl17}.
In this paper, we further connect the dynamics of vortices to the properties of the singularities in the Brillouin zone.
Especially, we compare the distribution of the momentum-resolved Loschmidt echo
and the number of vortex orbits between the accidental and the topological DQPTs in general two-band models.
Our results are complementary to previous ones.

The paper is organized as follows. Sec.~\ref{sec:dqpt} recalls the definition of DQPT and steady-state transition.
Sec.~\ref{sec:merge} shows why the DQPT merges into a steady-state transition
as the characteristic time diverges. Sec.~\ref{sec:scaling} discusses the scaling of the
dynamical free energy in the vicinity of the steady-state transition. Sec.~\ref{sec:dynamics}
is devoted to the dynamics of vortices. Sec.~\ref{sec:con} is a short summary.

\section{DQPTs and steady-state transitions}
\label{sec:dqpt}
We will illustrate the concepts of the DQPTs
and steady-state transitions by considering a two-dimensional Chern insulator. Its Hamiltonian
in momentum space is expressed as
$ \hat H = \sum_{\textbf{k}} \hat c^\dag_{\textbf{k}} \mathcal{H}_{\textbf{k}} \hat c_{\textbf{k}}$,
where $\hat c_{\textbf{k}} = \left(\hat c_{\textbf{k}1} ,\hat c_{\textbf{k} 2}\right)^T$ is the fermionic
operator and the sum of $\textbf{k}$ is over the Brillouin zone.
The two different components of fermions may refer to the
spin of electrons, the internal state of atoms or the sublattice
index of a complicated lattice.
The $2\times 2$ matrix $\mathcal{H}_{\textbf{k}}$ can be generally decomposed into
$\mathcal{H}_{\textbf{k}}=\vec{d}_{\textbf{k}} \cdot \vec{\sigma}$,
where $\vec{\sigma}=(\sigma_1,\sigma_2,\sigma_3)$ denotes the Pauli matrices
and $\vec{d}_{\textbf{k}}=\left( d_{1\textbf{k}},d_{2\textbf{k}},d_{3\textbf{k}}\right)$
is the coefficient vector.
This model has two energy bands with the energy $\pm d_{\textbf{k}}$, respectively, with
$d_{\textbf{k}}$ ($\geq 0$) the length of $\vec{d}_{\textbf{k}}$.
In the ground state, the negative-energy band
is fully occupied, but the positive-energy band is empty. The Hall conductance
in the ground state is well known to be
$\sigma_H = Ce^2/h$~\cite{shen}, where the Chern number $C$
is robust against a deformation of the Hamiltonian
except that the gap of the Hamiltonian closes and reopens accompanied by a discontinuous jump of $C$.

We denote the tunable gap parameter of $\mathcal{H}_{\textbf{k}}$ as $m$
whose absolute value is the energy gap.
And $m$ changes the sign as the gap closes and reopens.
The system is initially prepared in the ground state with the gap parameter being $m_i$.
To drive the system out of equilibrium, we suddenly change the gap parameter from $m_i$ to $m$.
Due to the integrability of the model, the system
cannot thermalize, instead, it will evolve into a steady state described by
the so-called generalized Gibbs ensemble (GGE)~\cite{rigol07}. The Hall conductance
of this nonequilibrium steady state was estimated as a function of $m$, i.e.,
the gap parameter of the post-quench Hamiltonian.
This function is nonanalytic at $m = 0$ and its derivative diverges
in a logarithmic way as~\cite{Wang2016a}
\begin{equation}\label{eq:central}
\frac{d \sigma_{H}}{dm} \sim
 \frac{e^2}{h} \frac{\Delta C}{2\left|m_i \right|} \ln \left|m\right|,
\end{equation}
where $\Delta C$ is the change of $C$ at $m=0$.

The nonanalyticity of the Hall conductance indicates a steady-state transition at $m=0$.
In a topological quantum phase transition, the ground
state of a Hamiltonian changes its topology as the energy gap closes.
The above steady state after a quench is not a ground state, nor a thermal state.
The winding number of the Green's function describes the topology of this steady state, which has a jump
as the energy gap of the post-quench Hamiltonian closes~\cite{Wang2016a}. At the same time,
the observable on this steady state (the Hall conductance) displays a nonanalytic behavior.
The nonanalyticity in the observable and the change of the steady-state topology
define a steady-state transition, which is distinguished from the topological quantum phase transition
by the exotic scaling behavior of the Hall conductance (see Eq.~(\ref{eq:central})).

Eq.~(\ref{eq:central}) is obtained by using the fact that the scaling of $ \sigma_{H}$ in the limit $m\to 0$
is independent of the form of $\vec{d}_{\textbf{k}}$
but depends only upon its lowest-order expansion around some singularities $\textbf{q}$ in the Brillouin zone
with $\textbf{q}$ defined by $d_{\textbf{q}}(m=0)=0$~\cite{Wang2016a}.
Around each singularity $\textbf{q}$, $\vec{d}_{\textbf{k}}(m)$ is expanded into the Taylor series:
\begin{equation}\label{eq:expdvector}
\begin{split}
d_{1\textbf{k}} = & a_{1x} \Delta k_x + a_{1y} \Delta k_y + \mathcal{O}(\Delta k^2), \\
d_{2\textbf{k}} = & a_{2x} \Delta k_x + a_{2y} \Delta k_y + \mathcal{O}(\Delta k^2), \\
d_{3\textbf{k}} = & m + \mathcal{O}(\Delta k^2),
\end{split}
\end{equation}
where $\Delta \textbf{k}=\textbf{k}-\textbf{q}$, and $a_{jx}$ and $a_{jy}$ are the expansion coefficients.
Due to the conic structure of the spectrum, the energy gap
nearby $\textbf{q}$ is $2d_{\textbf{q}}(m)=2| m|$. Substituting the expansion of $\vec{d}_{\textbf{k}}(m)$
into the expression of the Hall conductance leads to Eq.~(\ref{eq:central}).
Especially, the change of the Chern number is expressed as
\begin{equation}\label{eq:chernnumberchange}
\Delta C= \textbf{sgn} \left(a_{2x}a_{1y}-a_{1x}a_{2y}\right).
\end{equation}

Besides the steady-state transition,
the Chern insulator also exhibits DQPTs under an appropriate choice of $m_i$ and $m$.
With $\ket{\Psi(0)}$ denoting the ground state of $\hat H(m_i)$,
the Loschmidt echo is defined as $\mathcal{L}(t)=\bra{\Psi(0)}e^{-i\hat H(m)t}
\ket{\Psi(0)}$. Similar to the free energy in equilibrium states, one
can define a dynamical free energy as $g(t)=-\ln\left(\left|\mathcal{L}\right|\right)/S$
with $S$ being the system's area. Without considering the interaction between particles,
$\mathcal{L}(t)$ is a product of echoes at different momentum, and then
$g(t)$ in the thermodynamic limit $S\to \infty$ can be expressed as~\cite{dora15}
\begin{equation}\label{eq:regtwhole}
\begin{split}
 g(t) = & -\frac{1}{8 \pi^2} \int d\textbf{k}^2 \ln \bigg[
 \cos^2\left(t d_{\textbf{k}}(m) \right) \\ & + \sin^2\left(t d_{\textbf{k}}(m) \right)
 \left(\frac{\vec{d}_{\textbf{k}}(m_i) \cdot \vec{d}_{\textbf{k}}(m)}{{d}_{\textbf{k}}(m_i)
 {d}_{\textbf{k}}(m) } \right)^2 \bigg].
 \end{split}
\end{equation}
$g(t)$ becomes nonanalytic at some critical times $t^*$. This phenomenon is dubbed the DQPT.

\section{ Merging the DQPT into the steady-state transition}
\label{sec:merge}

The DQPT and the steady-state transition have indeed an intimate relation.
It was already known that the critical time when a DQPT happens is inversely proportional to
the energy difference between the upper and lower bands at the critical momenta where the
initial state is an equal-weight superposition of the post-quench eigenstates~\cite{Heyl2013a,Budich15}.
While the steady-state transition happens as the energy gap closes,
where the energy gap denotes the minimum of the energy difference between two bands over the Brillouin zone.
By using the singularity analysis, we find the relation between the  energy difference at the critical momenta
and the energy gap, and then obtain a model-independent expression of the critical time in terms of the latter.

Note that Eq.~(\ref{eq:regtwhole}) is an integral over the Brillouin zone.
In the integrand, the argument of the logarithmic function becomes zero if
there exist momenta $\textbf{k}$ (critical momenta) satisfying
$\vec{d}_{\textbf{k}}(m_i) \cdot \vec{d}_{\textbf{k}}(m)=0$
at the time $t^*_n=\displaystyle\frac{(2n+1)\pi}{2d_{\textbf{k}}(m)}$ with $n$
an integer~\cite{Heyl2013a,Budich15}. As a result,
$g(t)$ is nonanalytic at $t=t^*_n$ which is indicative of DQPTs.
The least critical time $t^*_0=\displaystyle\frac{\pi}{2d_{\textbf{k}}(m)}$ at $n=0$
is the characteristic time of DQPTs, which is also the time
period between two successive DQPTs. $t^*_0$ goes to infinity
if and only if $d_{\textbf{k}}(m)$ goes to zero. Since the energy gap of $\mathcal{H}_{\textbf{k}}(m)$,
i.e. $2\left| m\right|$, by definition must be less than or equal to
$2d_{\textbf{k}}(m)$ for any $\textbf{k}$ in the Brillouin zone,
$d_{\textbf{k}}(m)$ vanishes only if $m$ goes to zero.
What we have understood is that a steady-state transition happens at $m = 0$. Therefore,
the DQPTs in the limit $t^*_0 \to\infty$ must merge into a steady-state transition.

For obtaining the relation between the characteristic time $t^\ast_0$
and the gap parameter $m$, we replace $\vec{d}_{\textbf{k}}$ by the expansion~(\ref{eq:expdvector})
in the equation $\vec{d}_{\textbf{k}}(m_i) \cdot \vec{d}_{\textbf{k}}(m)=0$, which becomes
\begin{equation}\label{eq:didfexp}
\begin{split}
\sum_{j=1}^2 \left( a_{jx} \Delta k_x + a_{jy}\Delta k_y \right)^2 = - m_i m .
\end{split}
\end{equation}
Eq.~(\ref{eq:didfexp}) has solutions if and only if $m_i$ and $m$ have different signs.
Because the energy spectrum has a conic structure,
the quadratic form $\sum_{j=1}^2 \left( a_{jx} \Delta k_x + a_{jy}\Delta k_y \right)^2$
must be positive-definite. Therefore, the roots $\textbf{k}$ of Eq.~(\ref{eq:didfexp}) are located on a
circle centered at the singularity $\textbf{q}$. In the limit $m \to 0$, this circle
shrinks to the singularity $\textbf{q}$,
validating the above replacement of $\vec{d}_{\textbf{k}}$ by its lowest-order expansion around $\textbf{q}$.
By using Eq.~(\ref{eq:expdvector}) and Eq.~(\ref{eq:didfexp}), we express the characteristic time as
\begin{equation}\label{eq:criticaltimegeneral}
 \begin{split}
  t^*_0 = \frac{\pi}{2\sqrt{ m( m- m_i)}},
 \end{split}
\end{equation}
where we have set $\hbar =1$.
Fig.~\ref{fig:primary} schematically displays the change of $t^\ast_0$ as a function of $m$.
As $m_i$ is negative, the DQPTs exist only for $m > 0$.
$m=0$ is both the steady-state transition point and the point where DQPTs cease to exist.
In the vicinity of $m=0$, the characteristic time of DQPTs
changes as a universal function of the gap parameters $m_i$
and $m$ (Eq.~(\ref{eq:criticaltimegeneral})), being independent of the detail of the model.
Notice that, for the higher-order terms of $\vec{d}_{\textbf{k}}$ to be neglected,
both $m$ and $m_i$ should be close to zero, i.e., $\left|m_i\right|$ is in the same order of $\left| m\right|$.
Under this condition, the terms $\mathcal{O}(\Delta k^2)$ in the expansion of $\vec{d}_{\textbf{k}}$
result in a correction of $\mathcal{O}(|m|^3)$ to $\vec{d}_{\textbf{k}}(m_i) \cdot \vec{d}_{\textbf{k}}(m)$
which can be neglected.

\begin{figure}
\centering
\includegraphics[width=0.9\linewidth]{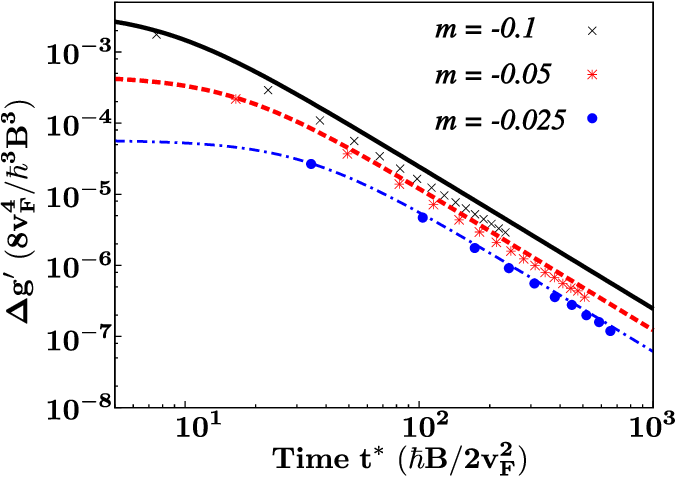}
\caption{(Color online)
$\Delta g' =\displaystyle \frac{dg}{dt}\bigg|_{t=t^{*+}}-\frac{dg}{dt}\bigg|_{t=t^{*-}}$
as a function of $t^*$ for the Dirac model (the dots) compared
with Eq.~(\ref{eq:cendptmain}) (the lines).
Different colors are for different values of $m$ and $m_i=-2 m$.
}
\label{fig:diffg}
\end{figure}

\section{The scaling of the dynamical free energy}
\label{sec:scaling}
The lowest-order expansion of $\vec{d}_{\textbf{k}}$ (Eq.~(\ref{eq:expdvector})) encodes all the information
for the steady-state transition at $m=0$. The higher-order terms are irrelevant to the scaling of $\sigma_H$,
reflecting the topological nature of the steady-state transition.
The expansion~(\ref{eq:expdvector}) also governs the
characteristics of the DQPTs in the vicinity of $m=0$.
This refers to not only the characteristic time but also the scaling of the dynamical free energy.
In the vicinity of $m=0$,
the derivative of $g(t)$ at the critical times satisfies (See App.~\ref{sec:appendix} for the derivation)
\begin{equation}\label{eq:cendptmain}
\begin{split}
& \frac{d g(t)}{dt}\bigg|_{t= t^{*+}} - \frac{d g(t)}{dt}\bigg|_{t= t^{*-}} \\
 & = \frac{1}{\left|a_{1x}a_{2y}-a_{2x}a_{1y}\right|} \times
 \\ & \frac{-m \left(m-m_i\right)^2}{2\sqrt{-m_i(m-m_i)}
\left(\displaystyle\frac{t^{*2}(m-m_i)}{4}-\frac{1}{m_i} \right)}.
 \end{split}
\end{equation}
If there are multiple singularities in the Brillouin zone, the right-hand side of Eq.~(\ref{eq:cendptmain})
should be a sum over all the singularities.
Note that the roots of $\vec{d}_{\textbf{k}}(m_i) \cdot \vec{d}_{\textbf{k}}(m)=0$
form a circle centered at $\textbf{q}$. If $d_{\textbf{k}}(m)$ is a constant on this circle,
$t^*=\displaystyle\frac{(2n+1)\pi}{2d_{\textbf{k}}(m)}$ for a specific $n$ is a point in the time axis,
and $t=t^{*\pm}$ in Eq.~(\ref{eq:cendptmain}) in fact means that $t$ approaches $t^*$ from above or from below, respectively.
If $d_{\textbf{k}}(m)$ on the circle $\vec{d}_{\textbf{k}}(m_i) \cdot \vec{d}_{\textbf{k}}(m)=0$ varies with $\textbf{k}$,
thereafter, $t^*$ becomes an interval (Fisher interval) in the time axis. In this case,
$t^{*+}$ and $t^{*-}$ denote the upper and lower endpoints
of the Fisher interval, respectively. In the limit $m \to 0$, a Fisher interval always narrows into a point,
the two critical times $t^{*\pm}$ then merge into one point
and $\displaystyle dg/dt |_{t=t^{*\pm}}$ becomes the one-sided limit of $dg/dt$.

Fig.~\ref{fig:diffg} shows the difference of $dg/dt$ at $t=t^{*\pm}$ for the Dirac model
which is defined by $\vec{d}_{\textbf{k}}=(\hbar v_F k_x, \hbar v_F k_y,mv^2_F - B \hbar^2 k^2)$
and the Brillouin zone being the infinite $k_x$-$k_y$ plane.
We set $\hbar=1$ and the Fermi velocity $v_F=1$ as the units, and
the irrelevant parameter $B$ is set to $2$ (i.e., $2/B$ is the unit of mass).
The expression of $\vec{d}_{\textbf{k}}$ is then simplified into
$\vec{d}_{\textbf{k}}=(k_x, k_y,m - 2 k^2)$.
$m$ is the gap parameter of the Dirac model.
The singularity in the Brillouin zone is at $\textbf{q}=0$, at which
one has $a_{1x}=a_{2y}=1$ and $a_{2x}=a_{1y}=0$.
We see that the numerical results fit well with Eq.~(\ref{eq:cendptmain}),
and the fit becomes even better for smaller $\left| m\right|$.

Eq.~(\ref{eq:cendptmain}) stands in the limit $m_i,m\to 0$
which corresponds to an infinitesimal quench crossing the
steady-state transition. Eq.~(\ref{eq:cendptmain}) reveals how the steady-state transition
controls the behavior of the dynamical free energy associated with DQPTs.
The difference of $dg/dt$ at $t=t^{*\pm}$ is independent of the detail of the model,
depending only upon $\left(a_{1x}a_{2y}-a_{2x}a_{1y}\right)$, $m_i$ and $m$
which are the characteristic parameters of the steady-state transition in the sense that
they determine the nonanalyticity of the Hall conductance
(see Eq.~(\ref{eq:central}) and~(\ref{eq:chernnumberchange})).

\begin{figure}
\centering
\includegraphics[width= 0.8\linewidth]{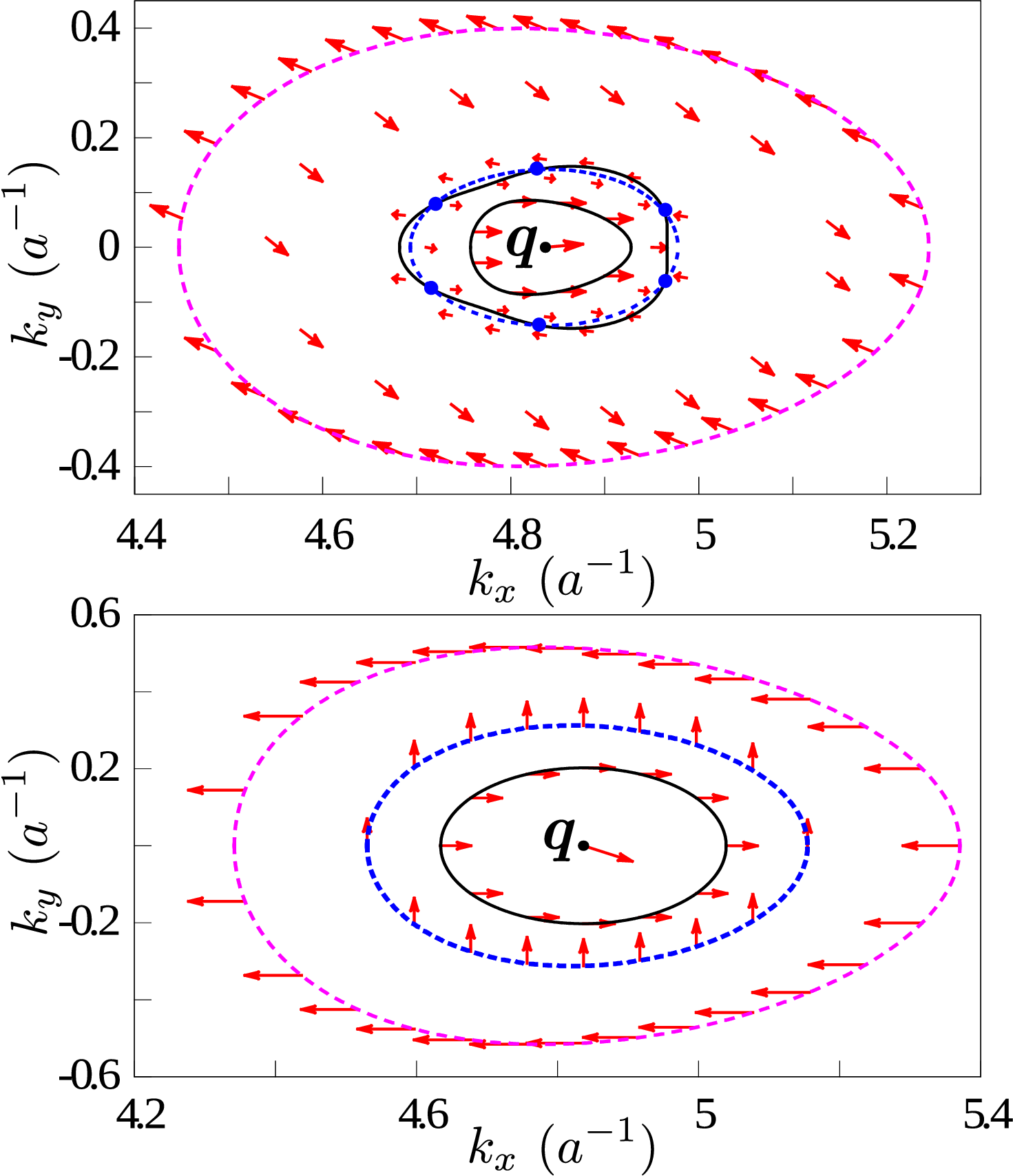}
\caption{The distribution of $\mathcal{L}_{\textbf{k}}$ (the red arrows)
around the singularity $\textbf{q}$
for the Haldane model at $m m_i >0$ (the top panel)
and at $mm_i <0$ (the bottom panel).
In the Haldane model, the edge of the honeycomb lattice $a$ is set to unity.
The solid black circles are the equi-occupation circles at $ m=-0.025$
and $ m_i=-0.475$ (the top panel) or at $ m=0.475$
and $ m_i=-0.475$ (the bottom panel), while
the dashed blue circles are the equi-energy circles at $d_\textbf{k}^2(m)=0.07$
(the top panel) or at $d_\textbf{k}^2(m)=2$ (the bottom panel). Their crosses, i.e. the
blue dots, are the vortices where $\mathcal{L}_{\textbf{k}}=0$.
The pink dashed circle connects the momenta far away
from $\textbf{q}$ where $\vec{d}_{\textbf{k}}(m)$ is almost parallel to $\vec{d}_{\textbf{k}}(m_i)$.
Another model parameter is $t'=4$.}
\label{fig:vortices}
\end{figure}

\section{Dynamics of the vortices}
\label{sec:dynamics}

A recent experiment~\cite{Flaeschner2016} measured the real-time dynamics of
the $\textbf{k}$-dependent Loschmidt echo $\mathcal{L}_{\textbf{k}}$,
i.e. the Loschmidt echo of a particle at the momentum $\textbf{k}$.
$\mathcal{L}_{\textbf{k}}$ is expressed as
\begin{equation}
 \mathcal{L}_{\textbf{k}} = \cos\left(t d_{\textbf{k}}(m) \right)
 + i \sin\left(t d_{\textbf{k}}(m) \right)
 \frac{\vec{d}_{\textbf{k}}(m_i) \cdot \vec{d}_{\textbf{k}}(m)}{{d}_{\textbf{k}}(m_i)
 {d}_{\textbf{k}}(m) }.
\end{equation}
It is related to the Loschmidt echo by $\mathcal{L} = \prod_{\textbf{k}} \mathcal{L}_{\textbf{k}}$.
The DQPTs are signaled by the zeros of $\mathcal{L}_{\textbf{k}}$ in the Brillouin zone.
A zero is also a vertex in the Brillouin zone,
by going around which $\mathcal{L}_{\textbf{k}}$ rotates by $360$ degrees in the complex plane.
We will analyze the dynamics of these vortices by using the
singularity $\textbf{q}$ and the expansion of $\vec{d}_{\textbf{k}}$ around it.
Our motivation is to obtain the general features in the dynamics of vortices
that are governed by the steady-state transition.

The vortices are obtained by solving $\mathcal{L}_{\textbf{k}}=0$
which is equivalent to the simultaneous equations $\vec{d}_{\textbf{k}}(m_i) \cdot \vec{d}_{\textbf{k}}(m)=0$
and $d_{\textbf{k}}(m)=\displaystyle\frac{\pi\left(2n+1\right)}{2t}$.
The roots of $\vec{d}_{\textbf{k}}(m_i) \cdot \vec{d}_{\textbf{k}}(m)=0$
form an equi-occupation circle surrounding the singularity $\textbf{q}$ in the Brillouin zone.
On the equi-occupation circle the asymptotic long-time occupations of the negative-energy and the positive-energy bands
are the same (both are $1/2$). The roots of $d_{\textbf{k}}(m)=\displaystyle\frac{\pi\left(2n+1\right)}{2t}$
form the equi-energy circles.
For a given time $t$, one can imagine a series of planes at the heights of
$\displaystyle\frac{\pi}{2t}, \displaystyle\frac{3\pi}{2t}, \displaystyle\frac{5\pi}{2t}, \cdots$ intersecting
the energy spectrum. Due to the conic structure of the spectrum,
these intersections are circles surrounding the singularity $\textbf{q}$ since $\textbf{q}$
is the minimum point of the spectrum.

The vortices are the crosses of the equi-occupation circle (the solid black lines in Fig.~\ref{fig:vortices})
and the equi-energy circle (the dashed blue lines in Fig.~\ref{fig:vortices}).
As $t$ increases, the planes $\displaystyle\frac{\pi}{2t}, \displaystyle\frac{3\pi}{2t}, \cdots$
moves downwards, thereafter,
the equi-energy circles shrink towards the singularity $\textbf{q}$.
A shrinking equi-energy circle will unavoidably meet the
equi-occupation circle surrounding $\textbf{q}$ and generate a family of
vortices (the blue dots in Fig.~\ref{fig:vortices})
at the time $t^{*-}_n$ which is the lower endpoint of a Fisher interval.
These vortices move on the equi-occupation circle and finally
annihilate each other at the time $t^{*+}_n$ (the upper endpoint of a Fisher interval),
before the equi-energy circle retracts into the equi-occupation circle.
Therefore, $\mathcal{L}_{\textbf{k}}$ exhibits vortices
if and only if there exist equi-occupation circles in the Brillouin zone.

In the discussion of DQPTs, two different cases must be distinguished.
In the case of $m m_i <0$,
i.e. a quench crossing the gap-closing point,
the existence of the equi-occupation circle and then
the DQPTs are topologically protected. One can obtain the solutions of
$\vec{d}_{\textbf{k}}(m_i) \cdot \vec{d}_{\textbf{k}}(m)=0$
by using the lowest-order expansion of $\vec{d}_{\textbf{k}}$ (Eq.~(\ref{eq:expdvector})).
On the other hand, as $m_i$ and $m$ have the same sign,
there is also possibility that the equi-occupation circles exist.
But the higher-order expansion of $\vec{d}_{\textbf{k}}$ must be considered
for obtaining the equi-occupation circles.
The DQPTs at $ m m_i >0$ are called the accidental DQPTs.

We employ the Haldane model~\cite{Haldane} as an example to demonstrate
the difference between the topological DQPTs and the accidental DQPTs.
The coefficient vector of the Haldane model
is $d_{1\textbf{k}}=\displaystyle\sum_{s=1}^3 \cos(\textbf{k}\cdot \textbf{a}_s)$,
$d_{2\textbf{k}}=\displaystyle\sum_{s=1}^3 \sin(\textbf{k}\cdot \textbf{a}_s)$, and
$d_{3\textbf{k}}=M-2t_2 \sin \phi \displaystyle\sum_{s=1}^3 \sin(\textbf{k}\cdot \textbf{b}_s)$,
where $M$ is a tunable parameter.
The gap-closing point of $M$ is at $M_c = 3\sqrt{3} t_2 \sin\phi $
with the corresponding singularity being $\textbf{q}=\left(8\pi/3\sqrt{3}, 0\right)$.
The energy gap parameter of the Haldane model is $m=M-M_c$
(see App.~\ref{sec:haldane} for more detail).
For the Haldane model, $d_{3\textbf{k}}$ in the neighborhood of $\textbf{q}$
can be expanded to the second order as $d_{3\textbf{k}}= m + \frac{9}{4}\sqrt{3}t_2 \sin\phi \Delta k^2 + \mathcal{O}(\Delta k^3)$.
At the same time, the lowest-order expansions of $d_{1\textbf{k}}$ and $d_{2\textbf{k}}$
are $d_{1\textbf{k}}=\frac{3}{2}\Delta k_x +\mathcal{O}(\Delta k^2)$ and
$d_{2\textbf{k}} = -\frac{3}{2}\Delta k_y+\mathcal{O}(\Delta k^2)$, respectively.
The solution of the equi-occupation equation now becomes
\begin{equation}\label{eq:acdqpt_k}
\begin{split}
 \Delta k^2  = & \frac{2}{9t'^2} \bigg( {-\left(1+t'\left(m_i + m\right)
 \right)}\\ & \pm {\sqrt{\left(1+t'\left( m_i + m\right)
 \right)^2-4 m_i m t'^2 }} \bigg),
 \end{split}
\end{equation}
where $t'=\sqrt{3}t_2\sin\phi$.
As $m_i m<0$, there always exists a single equal-occupation circle
(see Fig.~\ref{fig:vortices}, the bottom panel),
since the right-hand side of Eq.~(\ref{eq:acdqpt_k})
is larger than zero for either the sign ``+'' or the sign ``-''.
And in the limit $ m_i, m \to 0$, Eq.~(\ref{eq:acdqpt_k}) becomes
$\Delta k^2 \approx -4 m_i  m/9$ (using $\sqrt{1+x}\approx 1+x/2$),
which fits well with Eq.~(\ref{eq:didfexp}).
On the other hand, as $m_i m>0$,
the right-hand side of Eq.~(\ref{eq:acdqpt_k}) is larger than zero
for both ``+'' and ``-'' if $t'(m_i + m)<-1$,  but is always less than zero otherwise.
As $t'( m_i + m)<-1$, there simultaneously exist two equi-occupation circles surrounding $\textbf{q}$
(see Fig.~\ref{fig:vortices}, the top panel).
The DQPTs under the condition $m m_i >0$ and $t'( m_i + m)<-1$ are the accidental DQPTs.

In general, the number of equi-occupation circles surrounding the singularity
$\textbf{q}$ must be odd as $ m_i  m <0$, but even (including zero)
as $ m_i m >0$. This statement can be proved as follows.
At the singularity $\textbf{q}$, the coefficient vectors become
$\vec{d}_{\textbf{q}}(m_i)=\left(0,0, m_i\right)$ and $\vec{d}_{\textbf{q}}(m)=
\left(0,0, m\right)$.
As $m_i m>0$, $\vec{d}_{\textbf{q}}(m_i)$ and $\vec{d}_{\textbf{q}}(m)$ are in the same direction.
But they are in the opposite direction as $ m_i m<0$.
As $\textbf{k}$ moves in the Brillouin zone, both $\vec{d}_{\textbf{k}}(m_i)$ and $\vec{d}_{\textbf{k}}(m)$
rotate smoothly.
As $\textbf{k}$ is far away from $\textbf{q}$ (on the pink dashed circle in Fig.~\ref{fig:vortices}),
the contribution of $m_i$ ($m$) to the value of $d_{3\textbf{k}}(m_i)$
($ d_{3\textbf{k}}(m)$) can be neglected so that $d_{3\textbf{k}}(m_i)$
and $d_{3\textbf{k}}(m)$ are approximately the same and then $\vec{d}_{\textbf{k}}(m_i)$
and $\vec{d}_{\textbf{k}}(m)$ are in the same direction.
Note that we limit our discussion
in the vicinity of the steady-state transition, that is $\left|m_i\right|$ and $\left| m\right|$
are both small compared to the value of $\left|d_{3\textbf{k}}\right|$ far away from $\textbf{q}$.
Therefore, as $\textbf{k}$ moves from the singularity
to the pink dashed circle, it must cross the equi-occupation circles where
$\vec{d}_{\textbf{k}}(m_i) \perp \vec{d}_{\textbf{k}}(m)$ for even number of times
if $ m_i m>0$, but for odd number of times if $ m_i m<0$.

Fig.~\ref{fig:vortices} displays the distribution of $\mathcal{L}_{\textbf{k}}$ in the Brillouin zone.
On the pink dashed circle, $\vec{d}_{\textbf{k}}(m_i) \| \vec{d}_{\textbf{k}}(m)$
indicates that the positive-band occupation is $n_{\textbf{k}+}=0$
but the negative-band occupation is $n_{\textbf{k}-}=1$ and
$\mathcal{L}_{\textbf{k}}=e^{it d_{\textbf{k}}(m)}$. On the
equi-energy circles, the real part of $\mathcal{L}_{\textbf{k}}$
vanishes since $\cos(t d_{\textbf{k}}(m))=0$. On the
equi-occupation circles, $\vec{d}_{\textbf{k}}(m_i) \perp \vec{d}_{\textbf{k}}(m)$
indicates that the imaginary part of $\mathcal{L}_{\textbf{k}}$ vanishes.
The occupation at $\textbf{q}$ is normal ($n_{\textbf{q}+}=0$ and $n_{\textbf{q}-}=1$)
as $mm_i >0$ but it is reversed ($n_{\textbf{q}+}=1$ and $n_{\textbf{q}-}=0$) as $mm_i<0$.

Finally, the dynamics of vortices not only reflects the sign
of $m$, but also reflects the number of singularities in the Brillouin zone
if there exist multiple singularities.
For a generic model, if the energy gap closes simultaneously at multiple singularities,
these singularities are related to each other by a symmetry transformation.
An example is the Kitaev's honeycomb model which has two singularities
in the Brillouin zone~\cite{Kitaev06,Wang2016a}. As DQPTs happen, around each singularity,
a family of vortices are generated and annihilated.
The vortices surrounding a singularity transform together with the singularity under the symmetry transformation.
The number of vortex families is then equal to the number of singularities.
The latter is also equal to $\left|\Delta C\right|$ which
is the change of the Chern number at $m=0$. Because each singularity
contributes to $\Delta C$ by $\pm 1$ (see Eq.~(\ref{eq:chernnumberchange})) and the
contributions from different singularities are the same since they are related by a symmetry transformation.
Recall that $\Delta C$ plays an important role in determining the scaling
behavior of $\sigma_H$ at the steady-state transition (see Eq.~(\ref{eq:central})).
We then obtain another relation between the dynamics of vortices associated with DQPTs
and the scaling at the steady-state transition.

\section{Conclusions}
\label{sec:con}
We have shown that the DQPTs in a topological system always
merge into a steady-state transition driven by the closing and reopening of the energy gap.
By expanding the model Hamiltonian in the neighborhood of singularities in the Brillouin zone,
we explore the general properties of DQPTs in the limit of diverging characteristic time.
The characteristic time, the derivative of the dynamical free energy and the dynamics of vortices
associated with DQPTs display universal behavior
which are determined by the characteristic parameters at the steady-state transition.
Experimentally, the DQPT was observed in an optical lattice simulating
the Haldane model, where the energy gap can be tuned by the energy offset
between the $A$- and $B$-sublattice. It is then hopeful to observe the
universal behavior discussed in this paper.

\section*{Acknowledgement}
We would like to acknowledge the inspiring discussions with Markus Heyl
and his help in writing the paper. This work is
supported by NSF of China under Grant Nos.~11304280,~11372466 and~11774315.

\appendix

\section{Calculation of the dynamical free energy}
\label{sec:appendix}

To study the nonanalytic
behavior of the dynamical free energy at the critical times,
we notice that the dynamical free energy
is an integral of a logarithmic function over the Brillouin zone.
We divide the domain of integration
into the neighborhood of the singularity $\textbf{q}$ and the left area.
The neighborhood is large enough to cover the equi-occupation circle
$\vec{d}_{\textbf{k}}(m_i)\cdot \vec{d}_{\textbf{k}}(m)=0$.
The integral over the left area is an analytic function of time,
since the argument of the logarithmic function is nonzero once if $\textbf{k}$ is not
on the equi-occupation circle.
Therefore, the nonanalyticity of the dynamical free energy comes only from the integral over
the neighborhood of $\textbf{q}$. We define this integral as $g^{(\textbf{q})} (t)$, which is expressed as
\begin{equation}\label{eq:regtvecq}
\begin{split}
g^{(\textbf{q})} (t) = & -\frac{1}{8 \pi^2} \int_{{B}_\eta(\textbf{q})} d\textbf{k}^2 \ln \bigg[
 \cos^2\left(t d_{\textbf{k}}(m) \right) \\ & + \sin^2\left(t d_{\textbf{k}}(m) \right)
 \left(\frac{\vec{d}_{\textbf{k}}(m_i) \cdot \vec{d}_{\textbf{k}}(m)}{{d}_{\textbf{k}}(m_i)
 {d}_{\textbf{k}}(m) } \right)^2 \bigg],
 \end{split}
\end{equation}
where ${B}_\eta(\textbf{q})$ denotes the neighborhood of $\textbf{q}$ that covers the equi-occupation circle.

In the limit $m\to 0$, the equi-occupation circle shrinks to $\textbf{q}$,
thereafter, the neighborhood ${B}_\eta(\textbf{q})$ can be chosen to be arbitrarily small.
We can then substitute the lowest-order expansion of $\vec{d}_{\textbf{k}}$
into Eq.~(\ref{eq:regtvecq}) to calculate it.
We perform a linear transformation of coordinates in momentum space by making
$\sum_{j=1}^2 \left( a_{jx} \Delta k_x + a_{jy}\Delta k_y \right)^2 \to \Delta k^2$.
In the new coordinate system, the integrand has rotational symmetry.
And the equi-occupation circle is now
a circle of radius $\sqrt{-  m_i  m}$ centered at $\textbf{q}$.
Therefore, we choose ${B}_\eta(\textbf{q})$ to be a circle of radius $\sqrt{\eta}>\sqrt{-  m_i  m}$.
After we integrate out the azimuth angle, Eq.~(\ref{eq:regtvecq}) becomes
\begin{equation}\label{eq:intradial}
 \begin{split}
g^{(\textbf{q})} (t) = & -\frac{\displaystyle\int_0^\eta d\left(\Delta k^2\right)
\ln \left(G(\Delta k^2)\right)}{8 \pi|a_{1x}a_{2y}-a_{2x}a_{1y}|}
 \end{split}
\end{equation}
with
\begin{equation}\label{eq:momecho}
 \begin{split}
G(\Delta k^2) = &
 \cos^2\left(t \sqrt{m^2+\Delta k^2} \right) + \sin^2\left(t \sqrt{m^2+\Delta k^2} \right) \\ &
\times \frac{ \left( m_i m +\Delta k^2 \right)^2}{ \left( m^2_{i}+\Delta k^2\right)
 \left( m^2+\Delta k^2 \right) }.
 \end{split}
\end{equation}

In Eq.~(\ref{eq:intradial}) the integrand
has a singularity at $\Delta k^2 = - m_i m$ at which
$G(\Delta k^2)$ vanishes at the critical times
$t^* = \displaystyle\frac{\left(2n+1\right)\pi}{2\sqrt{m\left(m-m_i\right)}}$. We change the variable
of integration to $x=\Delta k^2 +m_i m$. The integral evaluates
\begin{equation}\label{eq:nonanagqt}
 g^{(\textbf{q})} (t) = \frac{1}{8 \pi|a_{1x}a_{2y}-a_{2x}a_{1y}|}
 \displaystyle\int_{m_i m}^{\eta+ m_i m} dx \frac{x G'(x)}{G(x)},
\end{equation}
where $\eta +m_i  m>0> m_i  m$ and
\begin{equation}
\begin{split}
G(x)=& \cos^2\left(t\sqrt{K^2+x}\right) \\ & +
\frac{x^2\sin^2\left(t\sqrt{K^2+x}\right)}
{\left(x+ m_i(m_i- m)\right)\left(x+K^2\right)}
\end{split}
\end{equation}
with $K=\sqrt{m( m- m_i)}$.
Note that we have neglected the analytic part in the expression of $g^{(\textbf{q})} (t)$.
$g^{(\textbf{q})} (t)$ in Eq.~(\ref{eq:nonanagqt}) is nonanalytic
at the critical times $t=t^*$.

Eq.~(\ref{eq:nonanagqt}) is still difficult to calculate.
But we are only interested in the nonanalytic behavior of $g^{(\textbf{q})} (t)$
at $t=t^*$. The nonanalyticity is independent of the domain of integration once if
the domain covers the singularity $x=0$ which corresponds to the equi-occupation circle.
Therefore, we choose the domain of integration
to be an infinitesimal neighborhood of $x=0$. In this neighborhood we can expand $G(x)$ into a power series as
\begin{equation}
 G(x) = \mu_0(t) + \mu_1(t) x + \mu_2(t) x^2 + \mathcal{O}(x^3).
\end{equation}
It is straightforward to verify $\mu_0(t^*)=\mu_1(t^*)=0$ but $\mu_2(t^*)>0$. As $t$ is close
enough to $t^*$, $\mu_2$ is always finite. We can then neglect the
higher-order terms $\mathcal{O}(x^3)$. Substituting the expansion of $G(x)$
into Eq.~(\ref{eq:nonanagqt}) and noticing $4\mu_0\mu_2-\mu_1^2>0$
as $t$ is close enough to $t^*$, we obtain
\begin{equation}
\begin{split}
  \int dx \frac{x G'(x)}{G(x)} =
  -\frac{\sqrt{4\mu_0 \mu_2-\mu_1^2}}{\mu_2}\tan^{-1}\left(\frac{2\mu_2 x+\mu_1}{\sqrt{4\mu_0 \mu_2-\mu_1^2}}\right).
\end{split}
\end{equation}
Here we only keep the nonanalytic part of the result.
By using the fact that the domain of integration
is an infinitesimal neighborhood of $x=0$, we can obtain the expression of
$g^{(\textbf{q})}(t)$ and then $dg^{(\textbf{q})}/dt$.
The nonanalytic behavior of the dynamical free energy can be expressed as
\begin{equation}\label{eq:appdgdtdiff}
\begin{split}
& \frac{d g(t)}{dt}\bigg|_{t= t^{*+}} - \frac{d g(t)}{dt}\bigg|_{t= t^{*-}} \\
 & = \frac{1}{\left|a_{1x}a_{2y}-a_{2x}a_{1y}\right|} \times
 \\ & \frac{- m \left( m- m_i\right)^2}{2\sqrt{- m_i( m- m_i)}
\left(\displaystyle\frac{t^{*2}( m- m_i)}{4}-\frac{1}{ m_i} \right)}.
 \end{split}
\end{equation}
Here $t=t^{*\pm}$ means that $t$ approaches $t^*$ from above or from below, respectively.

In the calculation we neglect the higher-order terms in the expansion of
$\vec{d}_{\textbf{k}}$. Because the higher-order terms are much smaller
compared to the lowest-order terms since we keep the domain of $\textbf{k}$
within an infinitesimal neighborhood of $\textbf{q}$.
The contributions from the higher-order terms
to $dg/dt$ can then be neglected in the limit $m, m_i \to 0$.
It is worth mentioning that the higher-order terms may also cause $d_{\textbf{k}}(m)$ varying
on the equi-occupation circle and then broaden the critical time $t^*$ into
a time interval (Fisher interval). In this case, $dg/dt$ becomes continuous at
$t^{*\pm}$ which denote the upper and lower endpoints of the Fisher interval, respectively,
and Eq.~(\ref{eq:appdgdtdiff}) then represents the difference of $dg/dt $ at the two endpoints.

\section{The Haldane model}
\label{sec:haldane}

The Haldane model describes the noninteracting fermions on a honeycomb lattice
which composes of two interpenetrating sublattices, i.e. the sublattice ``$A$'' and ``$B$''.
The model Hamiltonian includes the hopping term between the nearest neighbors
\begin{equation}
 \hat H_1 = \sum_{\langle \textbf{A}_i,\textbf{B}_j\rangle } \left(
 \hat c^\dag_{\textbf{A}_i} \hat c_{\textbf{B}_j} + \text{H.c.} \right),
\end{equation}
the hopping term between the next-nearest neighbors
\begin{equation}
\begin{split}
 \hat H_2 = & \sum_{\langle \langle \textbf{A}_i,\textbf{A}_j\rangle\rangle }
 \left( t_2e^{i\phi} \hat c^\dag_{\textbf{A}_i} \hat c_{\textbf{A}_j} + \text{H.c.} \right) \\
 & + \sum_{\langle \langle \textbf{B}_i,\textbf{B}_j\rangle\rangle }
 \left( t_2e^{i\phi} \hat c^\dag_{\textbf{B}_i} \hat c_{\textbf{B}_j} + \text{H.c.} \right),
 \end{split}
\end{equation}
and the onsite potentials breaking the inversion symmetry
\begin{equation}
 \hat H_3 = M \sum_{\textbf{A}_i}\hat c^\dag_{\textbf{A}_i} \hat c_{\textbf{A}_i}
 - M \sum_{\textbf{B}_i}\hat c^\dag_{\textbf{B}_i} \hat c_{\textbf{B}_i}.
\end{equation}
Here $\hat c^\dag_{\textbf{A}_i}$ and $\hat c_{\textbf{B}_j}$ are the fermionic operators,
$\textbf{A}_i$ and $\textbf{B}_j$ denote
different ``$A$'' and ``$B$'' sites, respectively, and $\langle \textbf{A}_i,\textbf{B}_j\rangle$
and $\langle \langle \textbf{A}_i,\textbf{A}_j\rangle\rangle$
denote the nearest-neighbor and the next-nearest neighbor relation, respectively.
$t_2$ is the hopping strength between next-nearest neighbors,
$\phi$ is the corresponding phase, and $M$ is the mass.

By using the Fourier transformation
$\hat c_{\textbf{k}1}= \sum_{\textbf{A}_j} \displaystyle\frac{e^{-i\textbf{k}\cdot \textbf{A}_j}}{\sqrt{L}}
\hat c_{\textbf{A}_j}$ and $\hat c_{\textbf{k}2}= \sum_{\textbf{B}_j} \displaystyle\frac{e^{-i\textbf{k}\cdot \textbf{B}_j}}{\sqrt{L}}
\hat c_{\textbf{B}_j}$ with $L$ being the total number of sites,
the Hamiltonian in momentum space becomes
$ \hat H = \sum_{\textbf{k}} \hat c^\dag_{\textbf{k}} \left(\vec{d}_{\textbf{k}} \cdot \vec{\sigma} \right)\hat c_{\textbf{k}}$
with $\hat c_{\textbf{k}} = \left(\hat c_{\textbf{k}1} ,\hat c_{\textbf{k} 2}\right)^T$.
The coefficient vector $\vec{d}_{\textbf{k}}$ can be expressed as
\begin{equation}\label{eq:vectordhaldane}
\begin{split}
d_{1\textbf{k}} = &\sum_{s=1,2,3} \cos\left(\textbf{k}\cdot \textbf{a}_s\right), \\
d_{2\textbf{k}} = & \sum_{s=1,2,3} \sin\left(\textbf{k}\cdot \textbf{a}_s\right), \\
d_{3\textbf{k}} =& M-2t_2 \sin\phi \sum_{s=1,2,3} \sin \left(\textbf{k}\cdot \textbf{b}_s\right).
\end{split}
 \end{equation}
Here we employ $6$ constant vectors
\begin{equation}
\begin{split}
	&\textbf a_1=\begin{pmatrix}0\\ -1\end{pmatrix}\ ,\quad
	\textbf a_2=\frac12\begin{pmatrix}\sqrt 3\\1\end{pmatrix}\ ,\quad
	\textbf a_3=\frac12\begin{pmatrix}-\sqrt3\\1\end{pmatrix}\ ,\\
	&\textbf b_1=\begin{pmatrix}\sqrt3\\0\end{pmatrix}\ ,\quad
	\textbf b_2=\frac12\begin{pmatrix}-\sqrt3\\3\end{pmatrix}\ ,\quad
	\textbf b_3=-\frac12\begin{pmatrix}\sqrt3\\3\end{pmatrix}\ .
\end{split}
\end{equation}
Note that the edge length of the honeycomb lattice is set to the unit of length.

The Haldane model has two gap-closing points which are at $M_c^{\pm} = \pm 3\sqrt{3} t_2\sin\phi$
with the corresponding singularities $\textbf{q}_+=\left( \displaystyle\frac{8\pi}{3\sqrt{3}}, 0 \right)$
and $\textbf{ q}_-=\left( \displaystyle\frac{4\pi}{3\sqrt{3}}, 0 \right)$. In this paper we focus
on $M_c^+$ and $\textbf{q}_+$. Around $\textbf{q}_+$ the coefficient
vector can be expanded into
\begin{equation}\label{eq:haldaneexpansion1}
\begin{split}
d_{1\textbf{k}}=&\frac{3}{2}\Delta k_x +\mathcal{O}(\Delta k^2),\\
d_{2\textbf{k}} =& -\frac{3}{2}\Delta k_y+\mathcal{O}(\Delta k^2),\\
d_{3\textbf{k}}=& \left(M-M^+_c\right) + \frac{9}{4}\sqrt{3}t_2 \sin\phi \Delta k^2 + \mathcal{O}(\Delta k^3).
\end{split}
\end{equation}


\bibliographystyle{apsrev}
\bibliography{literature}

\end{document}